# Front propagation in lattices with on-site bistable non-degenerate potential: multiplicity, bifurcations and route to chaos

I.B. Shiroky and O.V. Gendelman*

Faculty of Mechanical Engineering, Technion, Haifa 32000, Israel

* - contacting author, ovgend@technion.ac.il

*Propagation of transition fronts in models of coupled oscillators with non-degenerate on-site potential is usually considered in terms of travelling waves. We show that the system dynamics can be reformulated as an implicit map structure, and the travelling waves correspond to stable fixed points. Therefore, the loss of stability of such waves should follow well-known generic bifurcation scenarios. Then, one can expect a plethora of qualitatively different propagating-front solutions – multistable, multi-periodic, quasiperiodic and chaotic. All these solutions are readily revealed in a common $\varphi^4$ model.*

Lattice models of mobile defects and transition fronts (kinks) are ubiquitous in physics. The idea dates back to a seminal work of Frenkel and Kontorova (FK model), who used a discrete variant of the sine-Gordon equation to describe the motion of dislocations [1]. Atkinson and Cabrera [2] replaced the sinusoidal potential with a simplified piecewise-parabolic one to obtain analytically solvable equations for the original discrete lattice. A variation of the FK model with a triple-parabola potential was adopted for twinning dislocations in [3]. Similar model was further used to describe transition between dynamical phases, unlocking transitions and Aubry transitions in Josephson junction arrays [4]. In [5], Truskinovsky and Vainchtein used the model with piecewise linear inter-particle forcing to analyze martensitic phase transitions. Among other applications, one encounters detonation of primary explosives, domain walls in ferro-electrics [6], cracks in metals [7, 8], lattice distortion around twin boundaries [9], dry friction [10], statistical mechanics [11], crowdions in anisotropic crystals [12], motion of fronts in semiconductor superlattices [13], dynamics of carbon nano tubes foams [14], interaction of non-rigid walls in double-walled carbon nanotubes [15], superionic conductors [16], calcium release in cells [17] and others [18].

Propagating transition fronts in bi-stable medium are usually explored as travelling waves [2, 19, 20, 21, 5]. It is believed that in such models the front propagation velocity is uniquely defined by the system parameters. In linear chains with a piecewise parabolic on-site potential, above a certain threshold of the energetic gain at each site (or, equivalently, constant external forcing), the solution satisfies an "admissibility condition", meaning that at each instance the chain is separated into two segments, one is placed within the meta-stable well while the other within the stable one. Under this assumption, the model is solvable analytically, and, for velocities above a certain threshold, the solutions are indeed consistent with the aforementioned condition. For linear chains with a smooth non-degenerate bi-stable on-site potential, the analytical solution is not available. However, numerical studies and asymptotic analytical approximations indicate, in most cases, qualitatively similar behavior [20, 21, 19]. The stability of a driven topological soliton in a Frenkel-Kontorova model in the so called "fast region" has been studied by Braun et al [22]. In the underdamped case, the solution loses stability by appearance of a discrete breather within the tail. Recent studies devoted to the effect of nonlinear inter-particle interactions [23, 20], demonstrated that in these cases the propagation of the travelling wave fronts is dominated mostly by the nonlinearity, if the latter is strong enough. The travelling fronts reach very large energy concentration, far supersonic velocity, and primarily depend only on general shape characteristics of the on-site potential, rather than on its fine-details. Other studies [24, 25] which also address the case of generic coupling with on-site damping, propose a relation between the transported kinetic energy, the dissipation ratio and the velocity.

Much less is known about the propagation of transition fronts for the parameter region that does not allow the "admissibility conditions" to be satisfied. Most commonly, it happens for low values of the energetic gain, where the front moves relatively slowly, and its velocity falls below the maximum group velocity of the linear substructure of the chain [26]. For the Atkinson-Cabrera model, below a certain threshold, travelling waves do not exist in this case [3], [21], [2], [27], [19]. This conclusion motivated studies that tried to construct solutions in the "problematic" region. Vainchtein in [3] implements an analytical approach proposed by Flytzanis et al [28] to study a generalization of the Atkinson-Cabrera model. In this model, the two convex parabolas are connected by a concave one (a spinodal region) and form a continuously differentiable function. If the spinodal region is wide enough, solutions with more complex structure exist. Specifically, solutions with low velocities exist and emit waves in both directions. Similar results were also observed by Peyrard and Kruskal [29] and Earmme and Weiner [27]. Another interesting observation, obtained numerically, is that by implementing a fully nonlinear potential, the slow solutions are further stabilized. Paper [30] focuses on the degenerate case of [3], where the spinodal region is with zero width, or equivalently the Atkinson-Cabrera model. By applying a similar analytical approach, new solutions, which apparently complete the gap where the basic admissibility condition are violated, are constructed. However, numerical simulations indicate that these solutions are unstable. Therefore, it seems that the basic piecewise parabolic model cannot be applied to describe slow propagation of dislocations. In [31] a tri-linear nearest neighbor interaction is considered and is also solved by applying the technique of [28]. As the spinodal region is increased, a richer structure of solutions is found: solutions which emit waves of different frequencies in both directions, and a kinetic relation with several segments separated by velocity gaps.

There is no reason to think that non-existence of the stable travelling-wave solutions implies that the transition front cannot propagate through the lattice. In principle, it seems possible that the propagating front just have more complicated structure than the one forced by the travelling-wave ansatz. In some support of this idea, rich patterns of possible front propagation were observed in continuous models of detonation waves based on the Burgers equation. Paper [32] proposes an extension of Fickett model [33] to describe chemical reaction with an induction zone followed by heat release zone. These models yield pulsating and even chaotic propagating solutions. In [34], a model that predicts a shock wave in detonation in chemical mixtures is considered. Although this model is a very simple scalar first order PDE, it is rich enough to produce instability and chaos through the classical sequence of period doublings.

Current Letter is devoted to the exploration of similar propagation patterns in typical lattice models with a non-degenerate bi-stable on-site potential. The main idea is that the problem of the front propagation may be formulated in a form of a nonlinear map, and the travelling-wave solution corresponds to the fixed point of this map. When the parameters vary, such fixed point cannot just disappear – generically, it either collides with a similar unstable fixed point, or loses stability through generic and well-known bifurcation scenarios [35, 36]. Among these bifurcations, one encounters the period-doubling and Hopf bifurcations that, in terms of original lattice equations, will correspond to stable propagating fronts that do not obey to the traveling-wave ansatz. Moreover, one can expect that such a property is generic, at least for the smooth lattice models.

To illustrate the idea, we consider a traditional chain of oscillators with a smooth nearest-neighbor coupling potential $V(\varphi_n - \varphi_{n+1})$ subject to a smooth bi-stable on-site potential $U(\varphi_n)$. It is described by the following set of equations:

$$\ddot{\varphi}_n + V'(\varphi_n - \varphi_{n+1}) + V'(\varphi_n - \varphi_{n-1}) + \xi\dot{\varphi}_n = -U'(\varphi_n) \tag{1}$$

$\varphi_n$ is the displacement of the $n^{th}$ particle from the initial equilibrium state (meta-stable). $\xi$ is the linear on-site damping coefficient. The mass of each particle is set to unity.

Generic smooth on-site potential is characterized by the energetic effect $Q$, the height of the potential barrier $B$, the coordinate of the barrier $b$ and the coordinate of the stable state $\varphi^*$, as illustrated in Figure 1. The condition for non-degeneracy of the potential is $Q > 0$.

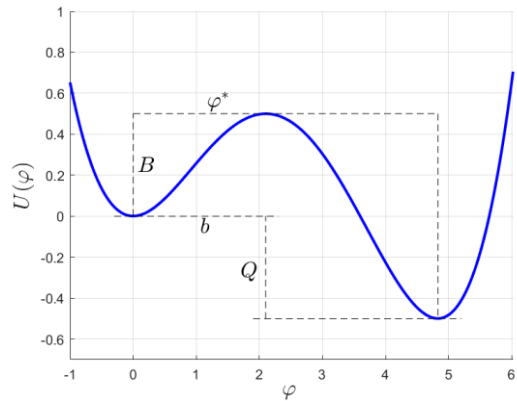

*Figure 1 - On-site bi-stable potential $U(\varphi)$*

We consider the propagation of a general transition front (not necessarily the travelling wave) and therefore assume that every particle passes in the first time from the metastable to the stable well in successive order. Each particle passes the barrier in the first time at certain time instance $t_k$:

$\varphi_k(t_k) = b$, $t_i < t_j \forall i < j$; $i, j \in \mathbb{Z}$. Without loss of generality we set $t_0 = 0$. The state vector is defined as follows:

$$\bar{X}(t) = \begin{pmatrix} \bar{\Phi} \\ \bar{V} \end{pmatrix}, \bar{\Phi} = \begin{pmatrix} \vdots \\ \varphi_{n-1}(t) \\ \varphi_n(t) \\ \varphi_{n+1}(t) \\ \vdots \end{pmatrix}, \bar{V} = \frac{d\bar{\Phi}}{dt} \qquad (2)$$

Then, if one denotes $\bar{X}_k = \bar{X}(t_k)$, then $\bar{X}_k$ constitutes a complete set of initial conditions for system (1) at time instance $t = t_k$. Therefore, it defines the state of the system when the next particle will cross the barrier in the first time. Formally, this obvious fact can be written down as a formal smooth map:

$$\bar{X}_{k+1} = \mathbf{F}(\bar{X}_k) \qquad (3)$$

Then, we define an invertible "shift backwards" operator acting on $\bar{X}(t)$ as follows:

$$\hat{S}(\bar{X}) = \begin{pmatrix} \bar{Q} \\ \bar{P} \end{pmatrix}, q_n = \varphi_{n+1}, \bar{P} = \frac{d\bar{Q}}{dt} \qquad (4)$$

Similar idea with introduction of cyclic shift matrix was used for stability analysis of the kink in driven Frenkel-Kontorova chain with periodic boundary conditions [22]. We introduce an auxiliary variable $\bar{Y}_k = \hat{S}^k \bar{X}_k$ and rewrite mapping (3) as follows:

$$\bar{Y}_{k+1} = \hat{S}^{k+1} \bar{F}(\hat{S}^{-k} \bar{Y}_k) \qquad (5)$$

Discrete map (5) is smooth due to the smoothness of system (1). Moreover, if System (1) has travelling-wave solutions in a form $\varphi_n(t) = \varphi(n - vt)$, then these solutions correspond to the fixed points of map (5). In this case $t_k = k/v$ and $v$ is the front velocity. Thus, as explained above, we expect that with variation of the system parameters these fixed points of the smooth map will lose stability with generic scenarios of co-dimension 1 and yield the stable propagating fronts of transition, which are not travelling waves.

To check this conclusion, we analyze a simple and very popular $\varphi^4$ model with linear nearest-neighbor coupling $V = (\varphi_n - \varphi_{n+1})^2 / 2$ and a quartic on-site potential $U(\varphi) = a_2 \varphi^2 + a_3 \varphi^3 + a_4 \varphi^4$ with coefficients adjusted to deliver the desired values of $Q, B, \varphi^*$. Bifurcation diagrams are produced for the control parameters $\varphi^*$ and $Q$. In the simulation, one of these parameters is varied slowly enough to ensure a quasi-static front propagation. In some instances, the parameter is varied also backwards, to explore the hysteretic behavior. The measured quantity is the velocity of transition of the front between two neighboring sites, defined as $V = V_j = (t_j - t_{j-1})^{-1}$. Here $j$ is the index of the last particle that switched to the stable domain at a given time instance. In the case of the travelling wave $V_j = v = \text{const}$ for all particles.

First, we present the bifurcation diagram of the front velocity as a function of parameter $\varphi^*$ (Figure 2).

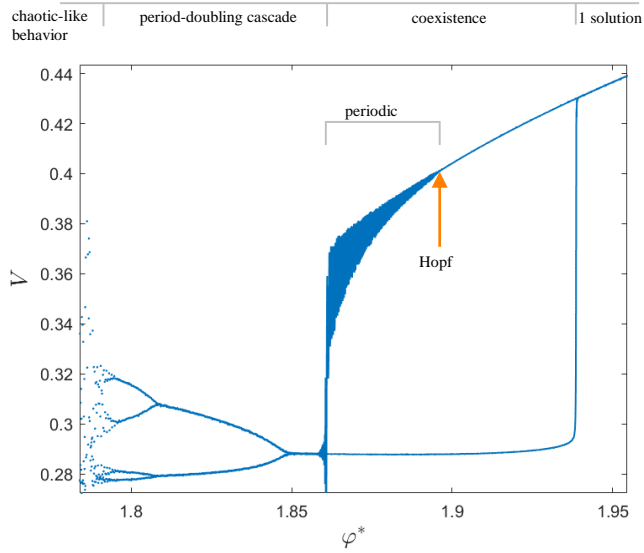

*Figure 2 – Bifurcation diagram of the front velocity as a function of $\varphi^*$ change. $B = 0.5$, $Q = 0.5$, $\xi = 0.02$*

One observes that for $\varphi^* > 1.94$ a single solution exists. The latter is similar to, for instance, the well-known travelling-wave fronts, which satisfy the "admissibility conditions" in the Atkinson-Cabrera model. In the $1.86 < \varphi^* < 1.94$ region, two different solutions coexist. One is a smooth continuation of the $\varphi^* > 1.94$ branch, while for the second, the front velocity is about 40% lower. The structure differences as well as the different rates of propagation of the two travelling-wave solutions are demonstrated in $n - t - \varphi$ space in Figure 3. Stability of both waves was verified numerically by introducing relatively strong random perturbations. The difference between the velocities is about 40%, and the frequency contents are considerably different. An additional interesting feature is that the upper branch undergoes a Hopf bifurcation at about $\varphi^* = 1.897$. Consequently, over the upper branch in the region $1.861 < \varphi^* < 1.897$, the propagating front violates the travelling-wave ansatz and propagates with oscillations. Example of such a behavior is presented in Figure 4. The amplitude of the front oscillations grows as $\varphi^*$ reaches the point $\varphi^* = 1.861$.

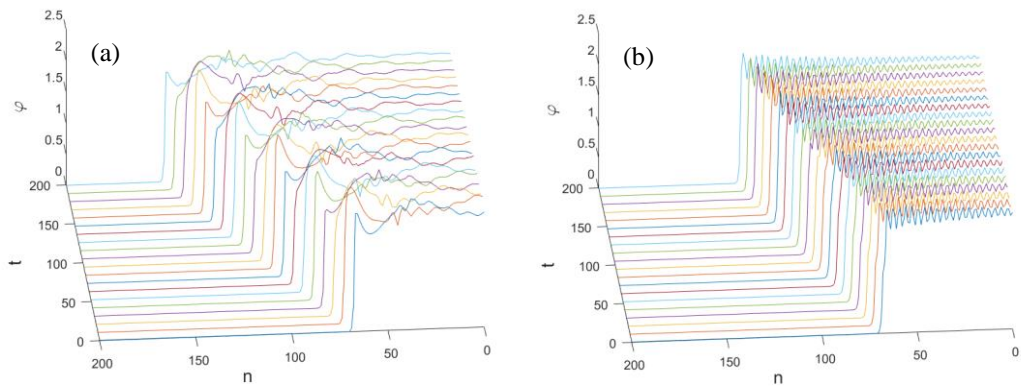

Figure 3 – Coexistence of propagating travelling waves for the same set of parameters; (a) $V = 0.404$, (b) $V = 0.2877$; Parameters: $\varphi^* = 1.9$, $Q = 0.5$, $B = 0.5$, $\xi = 0.02$

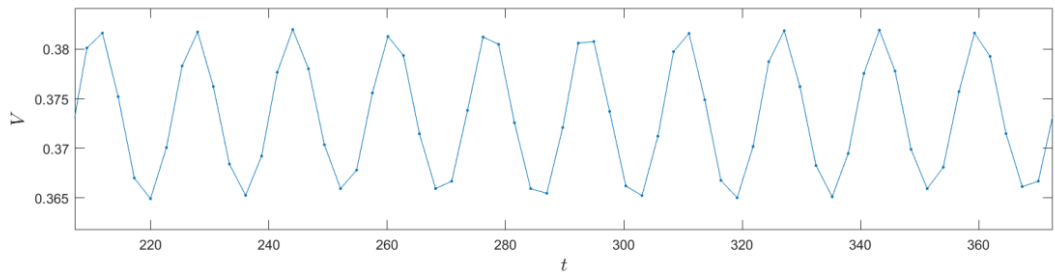

Figure 4 – An oscillatory front velocity following a Hopf bifurcation of the fast velocity branch; Parameters: $\varphi^* = 1.871$, $Q = 0.5$, $B = 0.5$, $\xi = 0.02$

As the value of $\varphi^*$ is further decreased, the front velocity enters a "period-doubling" cascade. The first doubling from a single branch to two branches occurs at $\varphi^* = 1.85$ and the lowest parameter value at which distinct branches of period doubling are observed is $\varphi^* = 1.791$ with eight branches. Within the doubling region the stationary propagation can be described in an averaged sense as follows:

$$\langle V \rangle = \frac{m}{\sum_{j=1}^{m} \Delta t_j} = \frac{m}{\sum_{j=1}^{m} \frac{1}{V_j}} \qquad (6)$$

Here $m$ is the number of branches of front velocity for the specific parameters $\varphi^*, Q$. The averaged front velocity is demonstrated in Figure 5.

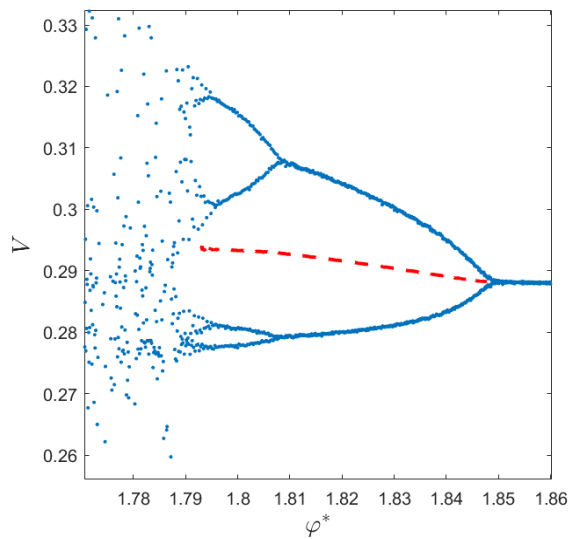

Figure 5 – Solution within the "period-doubling region"; Dashed red – averaged front velocity according to (6); Fixed parameters: $\xi = 0.02$, $Q = 0.5$, $B = 0.5$

Finally, the decrease in parameter $\varphi^*$ to the region $\varphi^* < 1.79$ leads to the chaotic-like behavior. Figure 6 presents an example of such a response. Subplot (a) shows that visually the response shape doesn't deviate much from typical responses in the non-chaotic regions. However, when exploring the plots of front location (b) and front velocity (c) as a function of time, the chaos is visually evident.

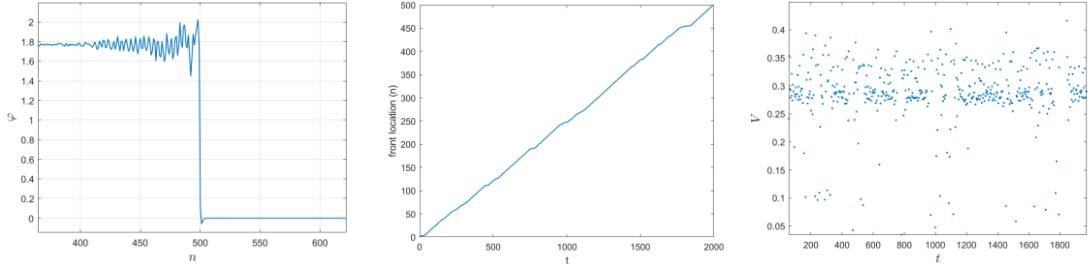

*Figure 6 – Wave propagation in the chaotic-like region;(a) $\varphi = \varphi(n)$; $t = 2000$ ,(b) front location as a function of time, (c) front velocity as a function of time; Parameters: $\varphi^* = 1.77$, $Q = 0.5, B = 0.5$, $\xi = 0.02$*

In Figure 7 the bifurcation diagrams are presented for a control parameter $Q$ for different fixed values of $\varphi^*$. At $\varphi^* = 1.84$ the period doubling occurs at $Q = 0.11$ and a period halving at $Q = 0.81$, hence this section doesn't lead to a chaotic like behavior. By taking a different section, $\varphi^* = 1.8$, as $Q$ is increased, the periods are doubled to the total number of 16, and at about $Q = 0.78$ the response is again chaotic-like.

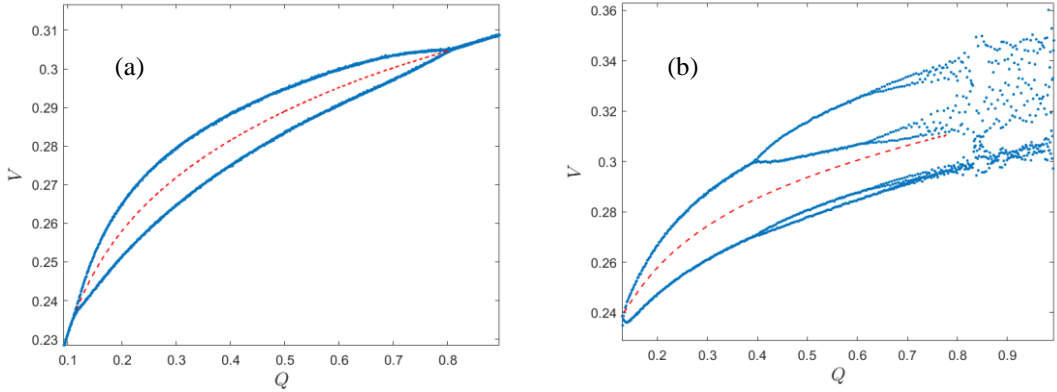

*Figure 7 – Bifurcation diagram of the front velocity as a function of Q change; (a) $\varphi^* = 1.84$, (b) $\varphi^* = 1.8$; Fixed parameters: $\xi = 0.02, B = 0.5$. Dashed red – averaged front velocity according to (6).*

To conclude, we reveal that regular co-dimension one bifurcations of the fixed points in an appropriately defined smooth map correspond to quite unusual patterns of the transition front propagation in the chains with a bistable non-degenerate on-site potential. One encounters quasiperiodic, multi-periodic and chaotic-like propagation – all these solutions are beyond the common travelling-wave behavior. These propagation regimes can be interpreted as the propagating kinks with coupled oscillatory states. It is well-known that, for instance, for the celebrated continuous degenerate $\varphi^4$ model one encounters oscillatory states around the stable kink solutions [37]. However, it seems that in the considered system the situation is different: the kinks with coupled periodic or chaotic oscillatory states can propagate through the lattice in the regions of the parametric space where the kinks without such coupled states do not exist or are unstable. Formation of discrete breather attached to the kink has been reported in driven Frenkel-Kontorova chain [22].

Another interesting finding is the co-existence of the travelling wave fronts with different velocities for the same set of parameters (Figure 3). For the nonlinear maps, this peculiarity amounts to a trivial possible co-existence of the stable fixed points. For propagating fronts, such behavior is not observed in common models, where the exact solutions are available. One can conjecture that these different regimes of propagation correspond to different possible phase lockings between the linear waves in the chain and the propagating front. Such multi-stability prevents existence of a single-valued "kinetic relation" for the front velocity as a function of the system parameters. Consequently, means that the mechanism of energy irradiation into the oscillatory tail cannot be mimicked by velocity-dependent "effective friction", at least in certain subset of the parametric space.

The last remark is that the observed regularities follow only from general properties of map (5). One can expect that the regimes of front propagation beyond the common travelling waves will be ubiquitous in similar lattice models, both in 1D and in higher dimensions.

The authors are very grateful to Israel Science Foundation (grant 1696/17) for financial support.